# Design approach for modern high-current, radio-frequency quadrupole accelerators with low emittance transfer

Chuan Zhang [1, #], Holger Podlech [2]

[1] *GSI Helmholtz Center for Heavy Ion Research, Planckstr. 1, Darmstadt, Germany*

[2] *Institute for Applied Physics, Goethe-University, Frankfurt a. M., Germany*

*Abstract*

Radio Frequency Quadrupole (RFQ) accelerators often need to face the challenges of space charge effects from high beam currents. This study investigated how to reach RFQ beam dynamics designs with not only high beam transmission and short structure length but also high beam quality simultaneously. To avoid beam instabilities induced by emittance transfer, a so-called $\frac{\varepsilon_l}{\varepsilon_t}$ (ratio of the longitudinal and transverse emittances) = 1.0 design guideline is being proposed. The application of this design guideline integrated with the BABBLE/NFSP design method for a high current, high frequency, high injection energy RFQ accelerator is described.

[#] e-mail: c.zhang@gsi.de

# I. INTRODUCTION

The R & D of a new generation of High Power Proton Accelerators (HPPA) with beam energies up to several GeV and power levels up to several MW has become a common interest for the particle accelerator community. Modern HPPA machines are of significant importance for promoting not only basic research but also advanced civil applications, because they can serve as spallation neutron sources, radioactive beam facilities, factories for other interesting secondary particles e.g. antiprotons and neutrinos, or accelerator-driven systems [1].

There are mainly three factors to increase the beam power: beam energy, beam duty factor, and beam current. Increasing the beam energy and the beam duty factor will bring most challenges to the accelerator technology development, while increasing the beam current can make the beam dynamics design more demanding, especially for the Radio Frequency Quadrupole (RFQ) accelerator. This is because the lower the beam energy is, the stronger the space charge effects are. As the typical injector to modern HPPA machines, therefore, the RFQ accelerator could be a bottleneck to the performance of the whole facility.

A worldwide survey of operational and planned hadron accelerators at the power frontiers including HPPA projects can be found in an overview plot from [2]. Some HPPA facilities accelerate protons directly, while some accelerate negative hydrogen ions in the linac part first and then strip two electrons from each ion at the injection into a synchrotron or an accumulator ring, which has the advantages to reduce the space charge effects at the injection, accumulate more particles, and finally lead to a higher beam brilliance. Apart from the intrabeam stripping issue, there is almost no difference in accelerating $H^-$ versus $H^+$ particles from the point of view of the linac beam dynamics.

TABLE I lists the main design parameters of some running high current $H^+$ or $H^-$ RFQ accelerators [3 – 6] realized for the HPPA projects (SNS, J-PARC, CSNS, and KOMAC) which appeared in the above-mentioned overview plot. Besides, two other similar machines, the LINAC4 RFQ [7] and the CPHS RFQ [8] are also included for a more comprehensive list. It can be seen that the typical frequency and output beam energy for the listed RFQ accelerators are 325 – 350MHz and 3MeV, respectively, except those parameters of the earlier SNS RFQ are a little different.

The purpose of this study is to develop a new design approach towards efficient RFQ accelerators of this type with high beam transmission, short structure length, and high beam quality. Therefore, a representative RFQ has been conceived, based on the following careful choices of its basic parameters:

- $f$ (frequency) = 325MHz: the lowest value of the listed RFQs for challenging the compactness as well as for having a higher current limit [9].

- $W_{out}$ (output beam energy) = 3MeV: the typical value of the listed RFQs.
- $I_{in}$ (design beam current) = 70mA: the highest value of the listed RFQs. A further requirement imposed on this study is that this RFQ should be able to work at currents up to 100mA, demonstrating its capacity to handle more serious space charge effects.
- $W_{in}$ (input beam energy) = 95keV and $\varepsilon_{in, trans., n., rms}$ (normalized rms transverse input emittance) = 0.3 $\pi$ mm mrad: as mentioned, the studied RFQ is supposed to be also feasible for a 100mA input beam, so relatively higher input energy to relax the space charge effects and relatively larger input emittance based on the state of the art ion source technology have been taken.
- $U$ (inter-vane voltage) = 80kV: moderate among the listed RFQs.

The design goals of this high current, high frequency, high injection energy RFQ (or HHH RFQ) are as the follows:

- The beam transmission should be ≥95% and ≥90% for the design current 70mA and the required higher current 100mA, respectively.
- The total structure length should be kept at ~3m, which is quite demanding, because the input energy of the HHH RFQ is almost double that of the other listed RFQs. It is known that for an adiabatic bunching section, the cell length is proportional to $\beta^3$, where $\beta$ is the ratio of the beam velocity to the speed of light in vacuum [10].
- High beam quality with not only small emittance growth but also as few as possible "unstable particles" should be reached. Here "unstable particles" mean the not-well-accelerated particles due to inadequate beam bunching and acceleration. Such particles will see wrong RF fields and can very likely be lost in later acceleration stages, so they are not favorable to meet the demands for modern accelerators e.g. hands-on maintenance, high reliability, etc. They are especially dangerous when superconducting (SC) cavities are employed directly after or close to the RFQ exit.

TABLE I. Design parameters of some realized ≥300MHz, 3MeV RFQs and design requirements for the HHH RFQ.

| | SNS | J-PARC RFQ-III | CSNS | KOMAC (PEFP) | CPHS | CERN LINAC4 | HHH |
|---|---|---|---|---|---|---|---|
| Ion | H⁻ | H⁻ | H⁻ | H⁺ | H⁺ | H⁻ | H⁺ |
| $f$ [MHz] | 402.5 | 324 | 324 | 350 | 325 | 352.2 | 325 |
| $W_{in}$ [keV] | 65 | 50 | 50 | 50 | 50 | 45 | 95 |
| $W_{out}$ [MeV] | 2.5 | 3.0 | 3.0 | 3.0 | 3.0 | 3.0 | 3.0 |
| $I_{in}$ [mA] | 60 | 60 | 40 | 20 | 50 | 70 | 70 – 100 |
| $U$ [kV] | 83 | 81 | 80 | 85 | 60 – 135 | 78 | 80 |
| $\varepsilon_{in, trans., n., rms}$ [π mm mrad] | 0.20 | 0.20 | 0.20 | 0.20 | 0.20 | 0.25 | 0.30 |
| $L$ [m] | 3.7 | 3.6 | 3.6 | 3.2 | 3.0 | 3.0 | ~3.0 |
| $T$ [%] | >90 | 98.5 | 97.1 | 98.3 | 97.2 | 95.0 | ≥95 at 70mA ≥90 at 100mA |

## II. COLLECTIVE INSTABILITIES AND HOFMANN CHARTS

To reach high beam transmission as well as high beam quality for a high current RFQ accelerator, one crucial issue is how to avoid beam instabilities induced by emittance transfer during the beam bunching and acceleration process [11, 12]. For beams with sufficient anisotropy of emittance and oscillation energy between different degrees of freedom, exchange-related resonances can occur and consequently cause beam instabilities, reduction of beam quality, and even beam losses [11, 12].

Systematic studies on collective instabilities for anisotropic beams have been done by I. Hofmann since several decades [11]. By imposing perturbations on an anisotropic Kapchinsky-Vladimirsky distribution in a constant focusing system with arbitrary focusing ratios and emittance ratios, calculations of the Vlasov equation in the two transverse dimensions were performed. I. Hofmann suggested that the same mechanisms of instability and similar thresholds could be applied for the longitudinal-transverse coupling resonances as well [12]. For different emittance ratios, the growth rates and thresholds of the resonances can be identified and visualized in the so-called "Hofmann Instability Charts" by two dimensionless parameters namely the tune ratio $\sigma_l/\sigma_t$ and the tune depression $\sigma/\sigma_0$, where $\frac{\sigma}{\sigma_0} = 1$ and $\frac{\sigma}{\sigma_0} \leq 0.4$ represent the zero-current case and the space-charge-dominated cases, respectively.

On the charts, the shaded areas mark the dangerous positions for emittance transfer, and the colors indicate the developing speed of these parametric resonances. Usually, the resonance peaks appear when $\frac{\sigma_l}{\sigma_t} = \frac{m}{n}$ ($m$ and $n$ are integers), e.g. at the positions: 1/1, 1/2, 2/3, etc. Actually, such resonances also exist at zero current because of the RF defocusing effect caused by the negative synchronous phase. However, the stop bands are significantly widened in the presence of strong space charge effects.

On the Hofmann Charts, the maximum spread of the safe tune depression (not shaded areas) always appears at a location where $\frac{\sigma_t}{\sigma_l} = \frac{\varepsilon_l}{\varepsilon_t}$ is satisfied and a resonance peak expected to be present is "killed". Because this condition can reach an energy-balanced beam and provide no free energy for driving the resonances, it is known as the equipartitioning (EP) condition [13]. Following this idea, a so-called "Equipartitioning Procedure" was developed and applied to the beam dynamics design of a 140mA, Continuous Wave (CW) deuteron RFQ [14]. The simulation results have shown that the EP design is successful and robust against very strong space charge effects as well as it needs very careful variations of the beam parameters to focus the tune footprints for the most part of the RFQ always on the EP line. For accelerating such a high current, CW $D^+$ beam to 5MeV safely, a very strict constraint on particle losses, especially those at over 1MeV, was imposed on the design, so it's worthy seeking top-level beam quality at the expense of a ~12.3m long RFQ (for the CDR Version; the RFQ was later shortened to ~8m in the Post-CDR phase) [14].

For RFQ designs, there is always a trade-off between beam quality, manufacturing complexity, investment, and other practical considerations. In most cases, a strict EP design is not necessary for RFQs which can tolerate a certain beam quality loss and some beam losses (e.g. the study presented in [15] has shown that relatively good beam quality is achievable by means of a relatively short RFQ if the tune footprints can cross the resonance peaks quickly enough). A tendency for modern large-scale linacs is to start applying the superconducting RF technology already in the very low energy part e.g. directly after or close to the RFQ exit, which requires the RFQ to deliver a high-quality beam not only for itself but also for a safe downstream operation. Therefore, a further investigation has been performed for trying to reach the EP-class beam quality without a strict EP design.

### III. DESIGN APPROACH FOR LOW EMITTANCE TRANSFER

This study is aiming to find useful hints from a series of Hofmann Charts for different emittance ratios (as shown in FIG. 1) and to propose an operable design approach for high current RFQs with high beam quality and low emittance transfer.

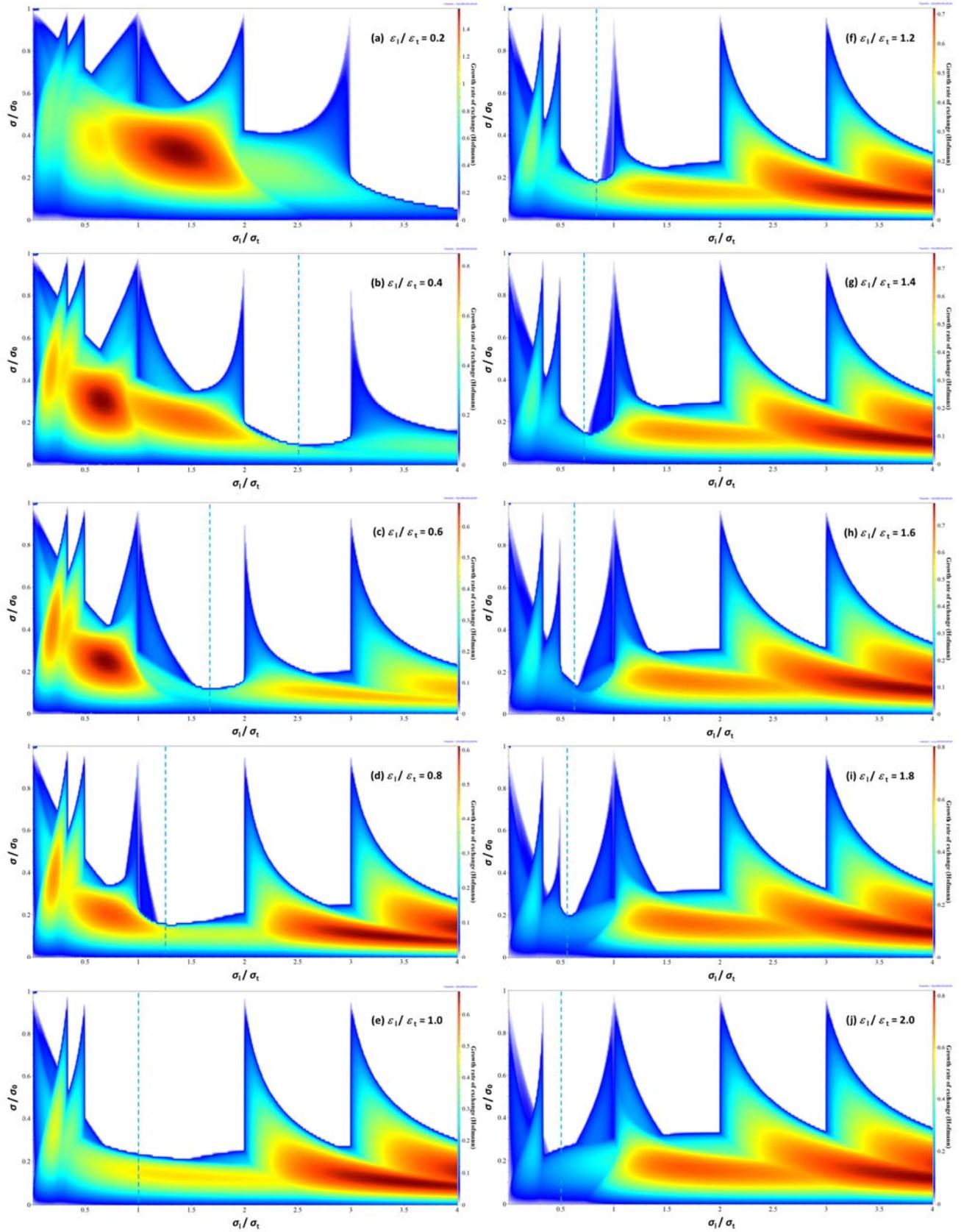

FIG. 1. Hofmann Charts for emittance ratios $\frac{\varepsilon_l}{\varepsilon_t}$ = 0.2 to 2.0, where the dashed lines are marking the EP positions (generated using the TraceWin Code [16]).

Generally speaking, after a beam is injected, the RFQ needs to firstly adapt the beam to a time-dependent focusing system by a short radial matching (RM) section, to then bunch it, and to finally accelerate it to the design energy. RFQ input beams have typically very small energy spread and very large phase spread, so for RFQ beam dynamics design studies it is usually assumed that the input energy spread and phase spread are 0 and ±180°, respectively, which is corresponding to a zero longitudinal emittance $\varepsilon_l$ at the RFQ entrance.

Along with the bunching and acceleration, $\varepsilon_l$ will be gradually increased to the final value. Besides high beam transmission and short structure length, another typical RFQ design goal is to hold the transverse emittance $\varepsilon_t$ constant throughout the accelerating channel and have as small as possible $\varepsilon_l$ at the exit, so this study is focusing on the Hofmann Charts for the emittance ratio $\frac{\varepsilon_l}{\varepsilon_t}$ up to 2.0 (see FIG. 1). The colors in the graphs indicate the growth rates of the resonances [12] (limited by the code, different graphs have slightly different colorbar ranges).

To maximize avoidance of the emittance transfer, the tune footprints of the beam should be kept in the "clean" area on the Hofmann Charts to the greatest extent. If it is inevitable to step into the resonance peaks, one should choose the low-growth-rate area and try to keep the trip as short as possible. FIG. 1 shows that the $\frac{\varepsilon_l}{\varepsilon_t}$ = 1.0 Hofmann Chart i.e. Graph (e) can provide the best working area for the tune footprints due to the following reasons:

- It allows safe evolutions of the beam parameters within very wide ranges of tune ratio ($\frac{\sigma_l}{\sigma_t}$ = 0.5 – 2.0) and tune depression ($\frac{\sigma}{\sigma_0}$ = ~0.2 – 1.0), respectively. This quasi-rectangle safe area is ideal to cover the most critical part for space charge effects during the RFQ bunching.
- The upper limit of its colorbar range is one of the lowest among all those shown, which means that its overall growth rate is relatively low.
- There are sufficient safety margins if the $\frac{\varepsilon_l}{\varepsilon_t}$ =1.0 condition is broken in a real machine due to different reasons e.g. errors. The boundary of the best working area can be more or less kept when $\frac{\varepsilon_l}{\varepsilon_t}$ is increased from 1.0 to 2.0 or decreased from 1.0 to 0.8, except the 1/1 resonance peak will appear. The larger the deviation is, the wider the 1/1 resonance peak is. For $\frac{\varepsilon_l}{\varepsilon_t}$ = 0.8 – 1.4, the 1/1 resonance peak is not so significant and its low growth rates (indicated by the blue colors) make it that this $\frac{\varepsilon_l}{\varepsilon_t}$ range is well acceptable for maintaining high beam quality.

- As the RFQ uses the same RF field for both bunching and acceleration, the longitudinal focusing force as well as the longitudinal phase advance will become smaller after the real acceleration starts. This usually moves the tune footprints into the $\frac{\sigma_l}{\sigma_t} \leq 0.5$ region and lets them enter the resonance peaks there. The caused emittance transfer can increase $\frac{\varepsilon_l}{\varepsilon_t}$ to be >1.0 (as the quickly increased beam velocity will weaken the transverse space charge effects naturally). However, all $\frac{\sigma_l}{\sigma_t} \leq 0.5$ resonance peaks in Graphs (f) – (j) of FIG. 1 have very low growth rates, so the beam quality will not drop much in case of a quick crossing.

To sum up, choosing $\frac{\varepsilon_l}{\varepsilon_t} = 1.0$ can be a useful guideline for designing RFQs with low emittance transfer.

For the RFQ beam dynamics design, the "LANL Four-Section Procedure" [17] is a classic technique. If the short RM section (typically only 4 – 6 cells long) at the entrance is ignored, it divides the main RFQ into three sequential sections: a "Shaper" (SH) section for prebunching, a "Gentle Buncher" (GB) section for main bunching, and finally an "Accelerator" (ACC) section for main acceleration.

The GB section holding the longitudinal small oscillation frequency of the particles and the geometric length of the separatrix constant is the key of this method for an adiabatic bunching (space charge forces have been neglected in the analysis for this section) [10]. In practice, all of the bunching cannot be done adiabatically in order to avoid a too long RFQ, so the SH section which ramps the phase and the acceleration efficiency linearly with axial distance for a fast prebunching is introduced. This is a potential source for unstable particles.

Another characteristic of the LANL method is that the mid-cell electrode aperture $r_0$ as well as the transverse focusing strength $B$ are held constant along the main RFQ. When the method was originally developed in 1978 – 1980, it was helpful for easing manufacturing and tuning (limited by the technologies at that time), but it is not reasonable for an RFQ in which the space change situation is changing.

To improve the LANL-style bunching process by adapting the transverse focusing strength to the changing space charge situation along the RFQ, a so-called "Balanced and Accelerated Beam Bunching at Low Energy" (BABBLE) method, which was previously known as "New Four-Section Procedure" (NFSP) [15, 18], has been developed. The BABBLE/NFSP method has enabled several efficient RFQ designs e.g. [15, 19] with both high beam transmission and short structure length, even at very high beam intensities e.g. 200 mA [18]. For the further improvement of the beam quality especially to minimize the unstable particles, a new design approach implementing the $\frac{\varepsilon_l}{\varepsilon_t}=1.0$ design guideline into the BABBLE/NFSP method is being proposed.

The new approach divides the main RFQ into also three sections: a main bunching (MB) section, a mixed-bunching-acceleration (MBA) section, and a main acceleration (MA) section in the following way:

- In the MB section, the beam goes through a so-called "$\varepsilon_l$ formation phase" in which the longitudinal emittance will be gradually increased from ~0 to a value that satisfies $\frac{\varepsilon_l}{\varepsilon_t} \cong 1.0$. In this phase, the tune ratio $\frac{\sigma_l}{\sigma_t}$ should be brought from 0 to a value between 0.5 and 2.0 as fast as possible, because there are some remarkable resonance peaks in the $\frac{\sigma_l}{\sigma_t} \leq 0.5$ region on the way to $\frac{\varepsilon_l}{\varepsilon_t} = 1.0$ (see Graphs (a) – (d) in FIG. 1). The phase advances $\sigma_l$ and $\sigma_t$ represent the longitudinal and transverse focusing strength, respectively. The LANL method holds the transverse focusing strength $B$ constant for the main RFQ, so usually $\sigma_t$ is already relatively large at the beginning of the bunching where $\sigma_l$ is still close to zero. This is not favourable to enter the safe $\frac{\sigma_l}{\sigma_t}$ range quickly. This problem can be overcome by the BABBLE/NFSP method, which applies a different strategy for this phase: 1) $B$ starts with a relatively small value because of the relatively weak transverse space charge effects before the bunching starts; 2) $B$ will be gradually increased to balance the increased longitudinal focusing during the bunching process; 3) longitudinally the synchronous phase $\varphi_s$ will be kept at ~ -90° to provide a maximum bunching and to increase $\sigma_l$ rapidly. In this way, a quick arrival of the tune footprints at the best working area provided by the $\frac{\varepsilon_l}{\varepsilon_t}=1.0$ Hofmann Chart is possible.

- The MBA section is the most critical part for beam bunching, because the space charge effects are getting stronger and stronger with the decreasing bunch size and become most significant at the end of this section. To avoid the emittance transfer maximally, this section should be well localized in the best working area of the $\frac{\varepsilon_l}{\varepsilon_t}=1.0$ Hofmann Chart. This will allow safe and fast beam parameter changes to hold $\frac{\sigma_l}{\sigma_t}$ staying in the range of 0.5 – 2.0. In this section, some acceleration is necessary for helping to relax the space charge effects.

- In the MA section, the longitudinal focusing force and the longitudinal phase advance will become smaller. If the transverse focusing strength is held as strong as with the LANL method, $\frac{\sigma_l}{\sigma_t}$ will decrease quickly. More reasonably, in the BABBLE/NFSP method $B$ is decreased such as to slow down the leaving speed of the tune trajectories from the best working area.

## IV. DESIGN & SIMULATION RESULTS OF THE HHH RFQ

The concrete beam dynamics design of the 70mA HHH RFQ has been achieved using the new design approach based on the combination of the BABBLE/NFSP method and the $\frac{\varepsilon_l}{\varepsilon_t}=1.0$ design guideline. The evolutions of the three main beam dynamics parameters i.e. electrode aperture $a_{min}$, electrode modulation $m$, and synchronous phase $\varphi_s$ along the designed HHH RFQ are shown in FIG. 2.

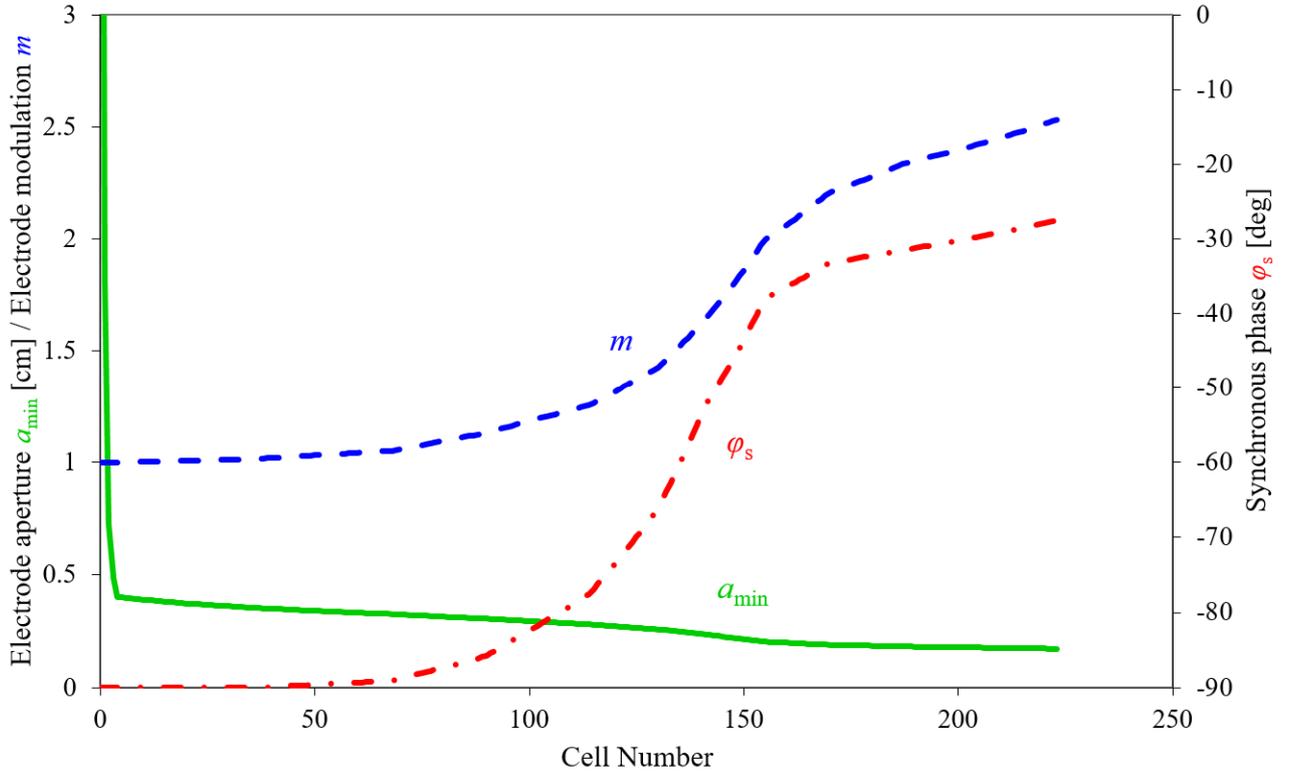

FIG. 2. Main beam dynamics design parameters of the RFQ.

The beam dynamics simulation was performed using the PARMTEQM code [20] with a 4D-Waterbag input distribution. FIG. 3 plots several key ratios between the longitudinal and transverse planes for our analyses, where $b$ and $a$ are the longitudinal and transverse rms beam sizes in mm, respectively. The partitions of the HHH RFQ can also be distinguished in the figure: 1) MB: Cell 0 to Cell 90; 2) MBA: Cell 91 to Cell 145; 3) MA: Cell 146 to the exit. At the end of the MB section,

$\frac{\varepsilon_l}{\varepsilon_t}$ reaches ~1.0 and the beam starts to enter the range of $\frac{\sigma_l}{\sigma_t} = 0.5 - 2.0$. Afterwards, the emittance ratio is kept at ~1.0 in the rest part of the RFQ, although $\frac{\sigma_l}{\sigma_t}$ drops to the ≤0.5 region in the last cells.

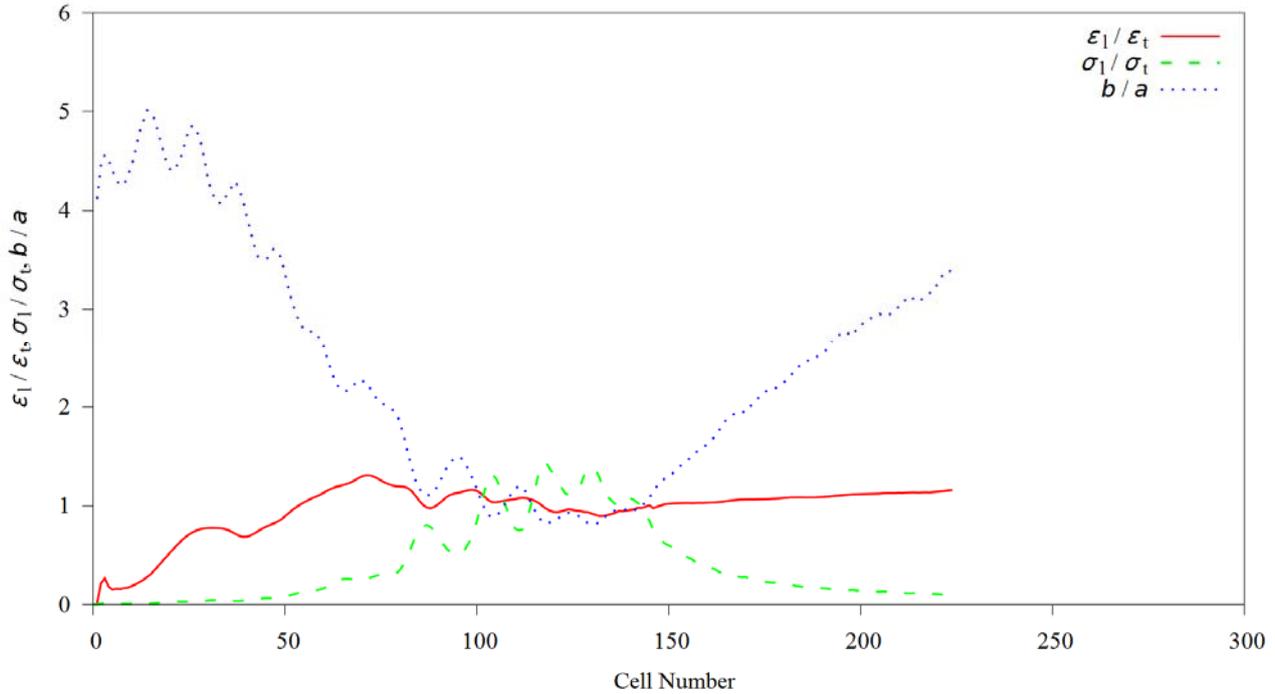

FIG. 3. Ratios of emittances $\frac{\varepsilon_l}{\varepsilon_t}$, phase advances $\frac{\sigma_l}{\sigma_t}$, and beam sizes $\frac{b}{a}$ along the RFQ.

The tune footprints of the HHH RFQ on the $\frac{\varepsilon_l}{\varepsilon_t} = 1.0$ Hofmann Chart are shown in FIG. 4. The different sections are marked in different colors i.e. red for MB, green for MBA, and blue for MA. Meanwhile, the solid and dashed curves represent the tune depressions in the transverse and longitudinal directions, respectively. It can be seen that the main part of the tune trajectories is located in the safe place on the chart so that this kind of bunching process can be performed much faster than that in the LANL-style GB section. This will consequently lead to a short structure length. In the MA section, the beam trajectories step into the instability region. On one side, the transverse tune depression $\frac{\sigma_t}{\sigma_{0t}}$ rises from ~0.65 to 0.8 as a result of weakened transverse space charge effects, and the corresponding tune trajectory travels still mostly in the resonance-free area, so the transverse emittance can stay essentially constant. On the other side, the longitudinal tune depression $\frac{\sigma_l}{\sigma_{0l}}$ drops from 0.5 to 0.3, but because the tune trajectory crosses the peaks shortly enough and the growth rates in this area are low enough, no obvious longitudinal emittance growth happened. Therefore, the $\frac{\varepsilon_l}{\varepsilon_t}$ is still held at ~1.0.

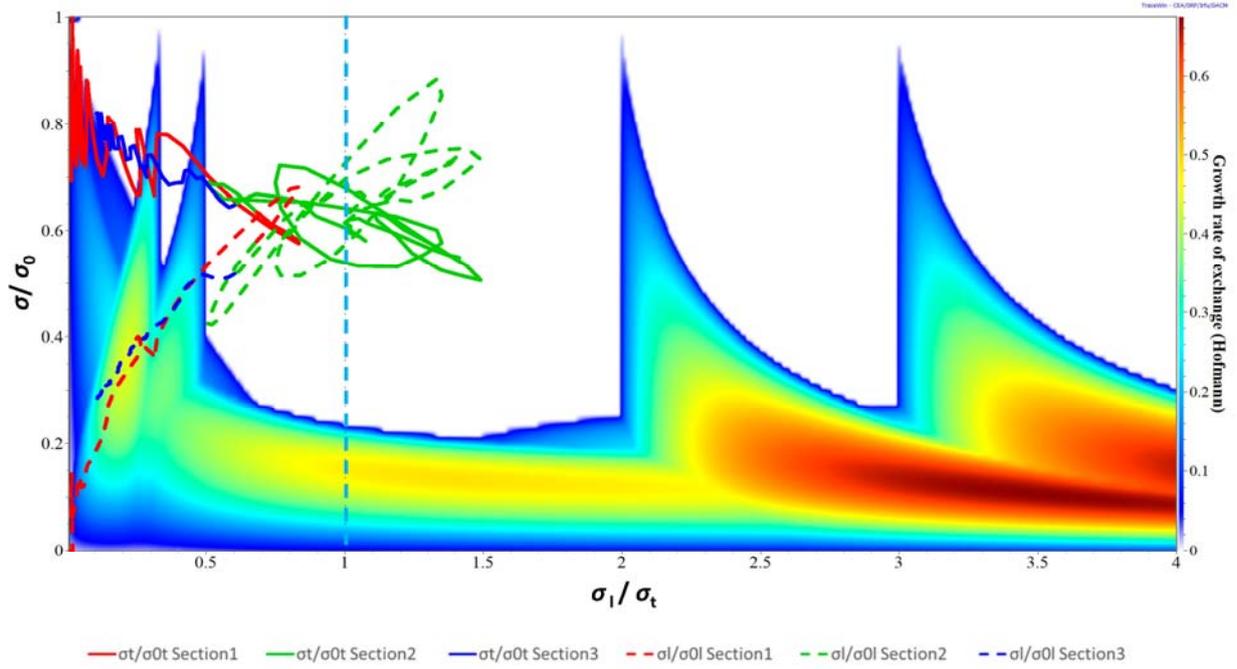

FIG. 4. Tune footprints of the RFQ at 70mA on the $\frac{\varepsilon_l}{\varepsilon_t} = 1.0$ Hofmann Chart.

The 70mA HHH RFQ design has been checked with different input transverse emittance values around the nominal one $\varepsilon_{in, trans., n., rms} = 0.3$ π mm mrad. FIG. 5 shows the evolution of the emittance ratio in the top graph (marked with the corresponding beam transmission efficiency $T$) as well as the evolution of the phase advance ratio in the bottom graph for each case, respectively.

The $\frac{\sigma_l}{\sigma_t}$ curves are still similar due to the same RFQ design, but the $\frac{\varepsilon_l}{\varepsilon_t}$ curves are not. Except the reference $\frac{\varepsilon_l}{\varepsilon_t}$ curve is held fairly stable around 1 for the main RFQ, all the others have some obvious oscillations, where the drops in the $\frac{\varepsilon_l}{\varepsilon_t}$ curves are reflecting particle losses in these cells. After the beam losses, however, the $\frac{\varepsilon_l}{\varepsilon_t}$ values will keep re-growing, which shows the instability of the beam.

When the input transverse emittance is increased from 0.1 to 0.5 π mm mrad, the resulting emittance ratio values at the end of MB (at Cell 90) are $\frac{\varepsilon_l}{\varepsilon_t} = $ 2.3, 1.3, 1.0, 0.9, and 0.8 respectively. Except for the 0.1 π mm mrad case, all final $\frac{\varepsilon_l}{\varepsilon_t}$ values at the end of the RFQ are still close to 1, which means that the oscillations have been caused by a few unstable particles and shows the tolerance of the design to the input emittance errors. It confirms also the acceptable $\frac{\varepsilon_l}{\varepsilon_t}$ range (0.8 – 1.4) mentioned in Section III. Of course, for $\varepsilon_{in, trans., n., rms} > 0.3$ π mm mrad, more beam losses occur, because a part of the beam is beyond the acceptance of the RFQ.

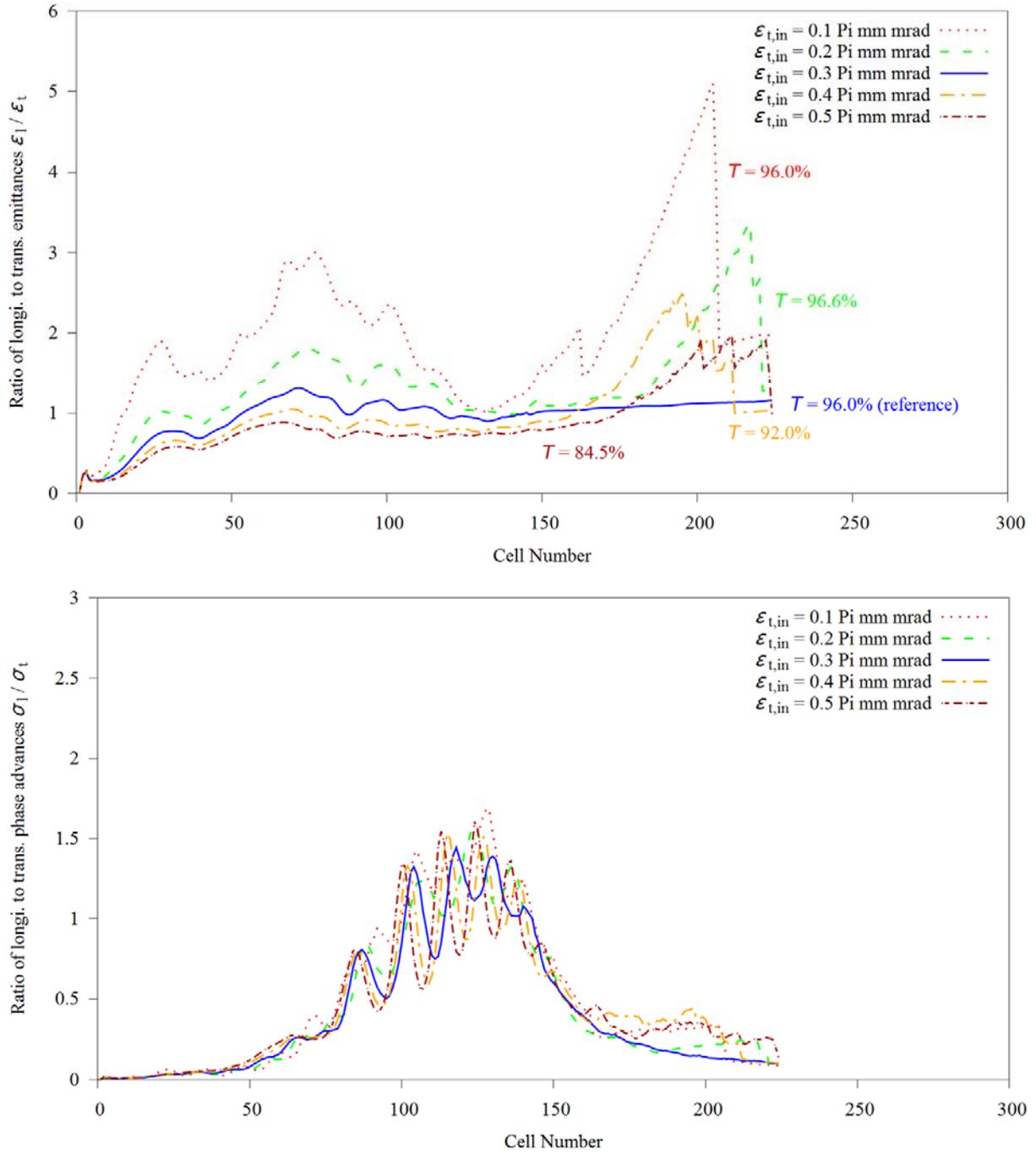

FIG. 5. Emittance ratio evolutions (top graph, marked with the transmission efficiency for each case) and phase advance ratio evolutions (bottom graph) along the RFQ at different input emittances.

As required, the HHH RFQ should also work at currents up to 100mA. FIG. 6 checks the tune footprints of the HHH RFQ at 100mA. Generally speaking, the behavior of the beam is still similar to that in the 70mA case. Of course, from 70mA to 100mA, the space charge effects are getting stronger, so the "center of gravity" of the beam trajectories has been lowered from ~0.65 to ~0.55 in the $\frac{\sigma}{\sigma_0}$ direction and the final longitudinal tune depression has fallen down from 0.3 to 0.2. But there is still sufficient "clean" space below the beam trajectories, which gives a hint that this RFQ can work for even higher currents.

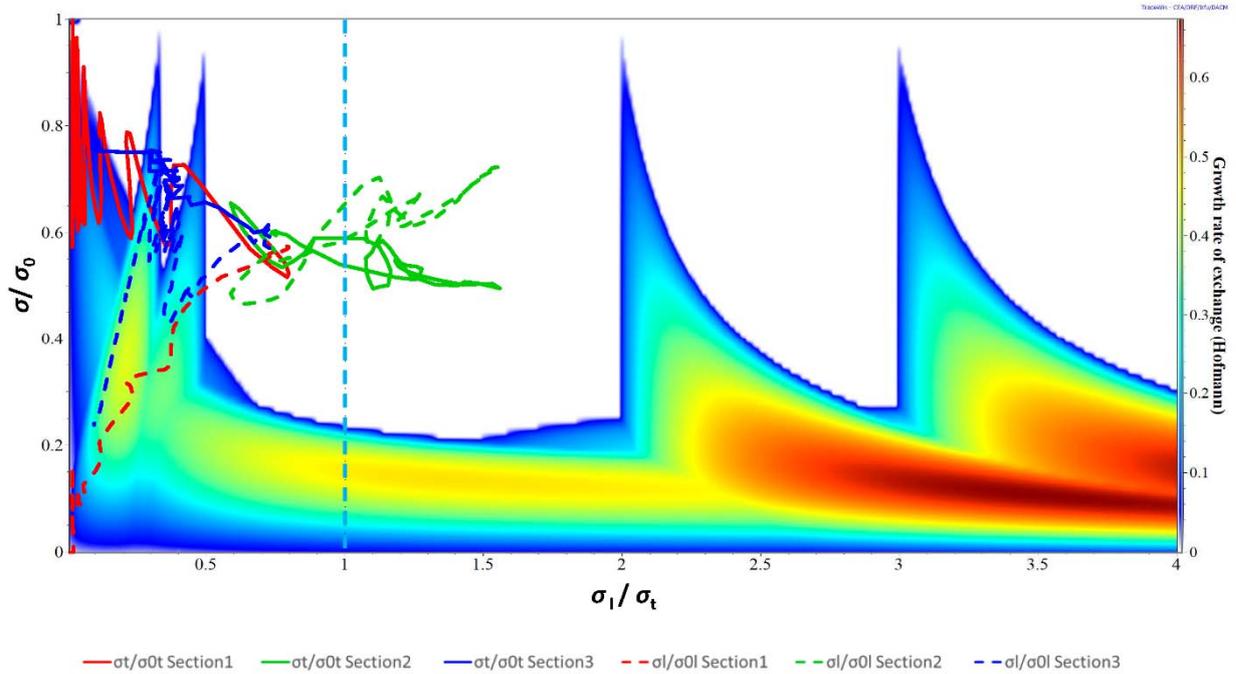

FIG. 6. Tune footprints of the RFQ at 100mA on the $\frac{\varepsilon_l}{\varepsilon_t} = 1.0$ Hofmann Chart.

Furthermore, the ratios of the longitudinal and transverse emittances as well as phase advances along the RFQ for both 70mA and 100mA are compared in FIG. 7. The evolutions of these parameters are fairly close to each other, except the emittance ratio of the 100mA beam has a few "jumps" in the main acceleration section due to the deviation to $\frac{\varepsilon_l}{\varepsilon_t} = 1.0$ at the MB end. From the curve of the emittance ratio calculated with the longitudinal emittance for 99% instead of 100% of the beam, it can be known that this difference has been caused by only ≤1% of halo particles.

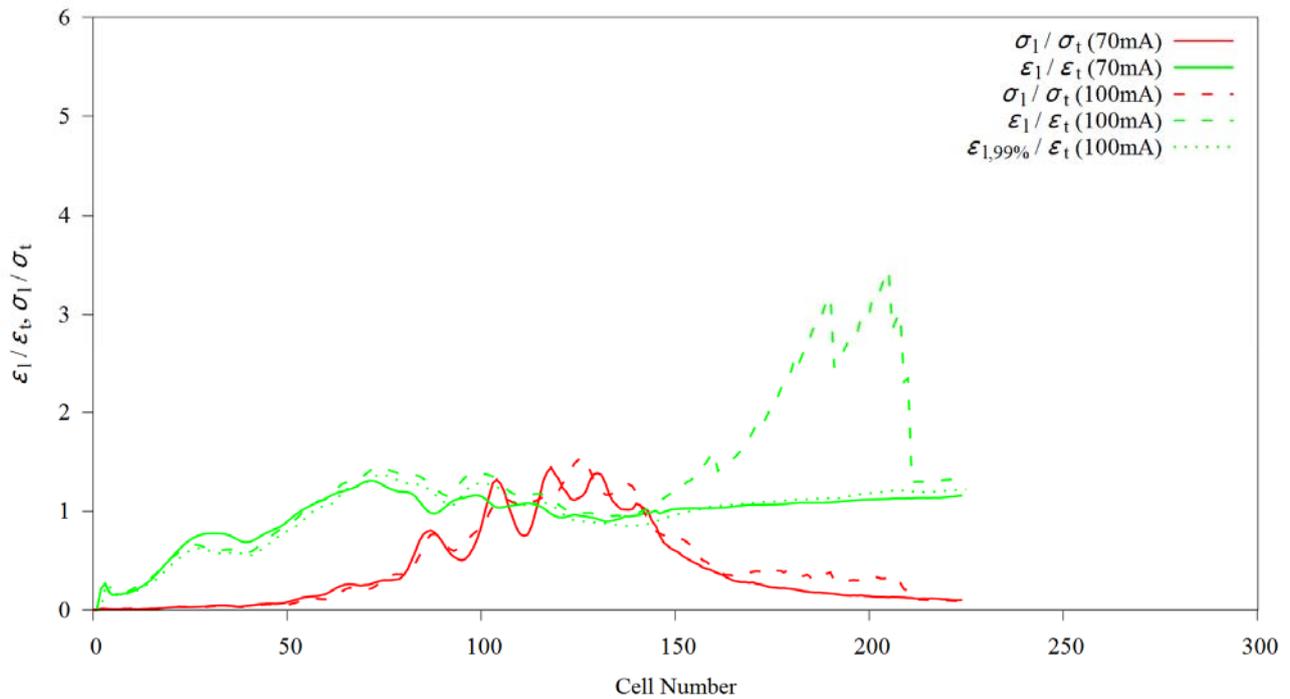

FIG. 7. Comparison of emittance ratios $\frac{\varepsilon_l}{\varepsilon_t}$ and phase advance ratios $\frac{\sigma_l}{\sigma_t}$ between 70mA and 100mA.

FIG. 8 shows the output particle distributions not only in the three phase spaces but also with the front, side and top views in the real space. Again, it shows the similarity between 70mA and 100mA, except the 100mA beam has slightly bigger beam sizes especially in the longitudinal direction. But in both cases, the beam is still well concentrated.

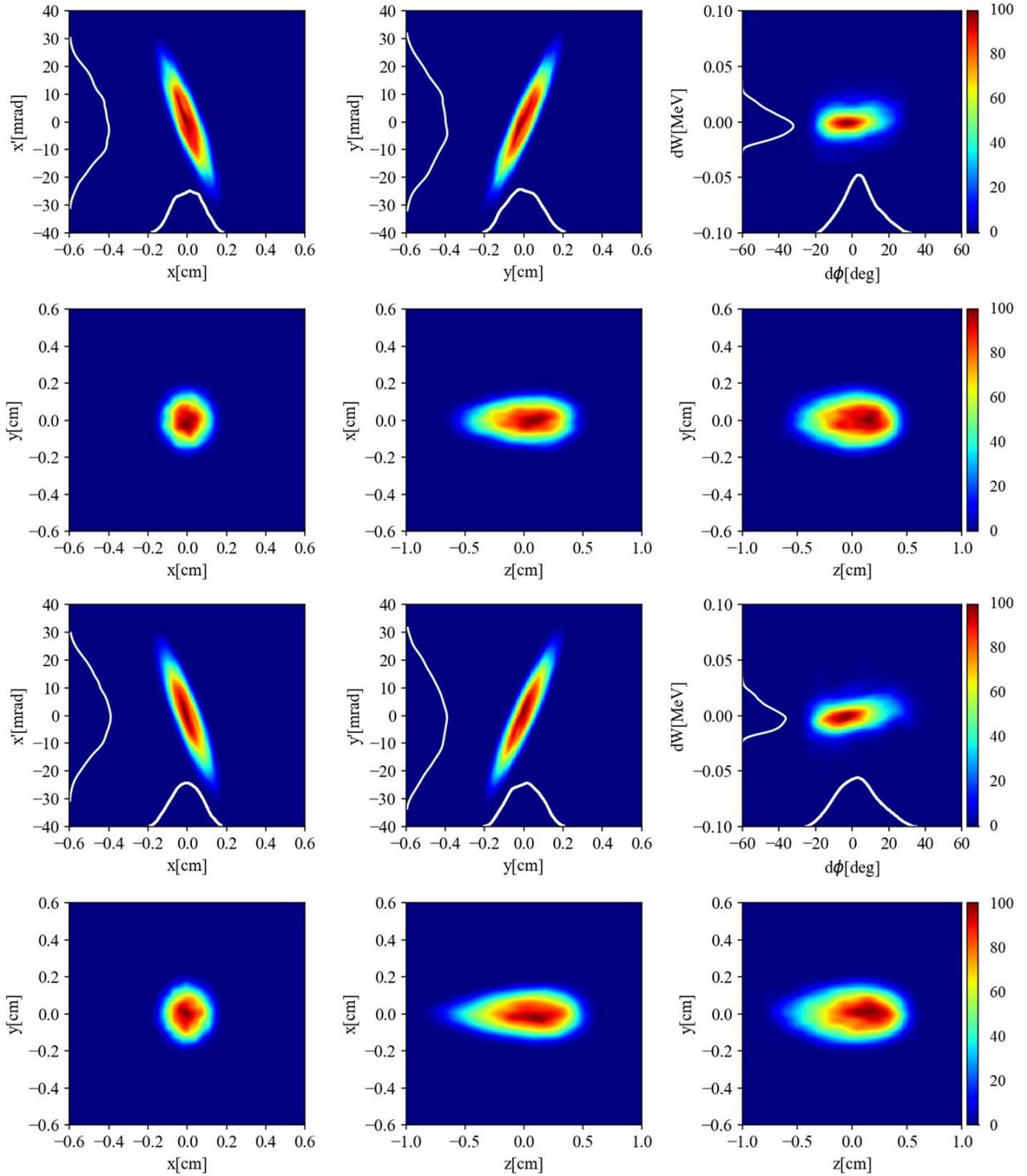

FIG. 8. Output particle distributions (top 6 graphs: 70mA; bottom 6 graphs: 100mA).

More detailed design and simulation results of the HHH RFQ at both 70mA and 100mA are given in TABLE II. The design goals with respect to the beam transmission have been well fulfilled. Although the design is not optimized for 100mA, *T* is still ≥90%.

Furthermore, the HHH RFQ has reached the shortest structure length i.e. 3m compared to the realized ≥300MHz, 3MeV RFQs listed in TABLE II. The efficiency of the new design approach can be seen from the fact that the HHH RFQ is 60cm – 70cm shorter than the SNS RFQ and the CSNS RFQ (both based on the traditional LANL method) as well as the J-PARC RFQ-III (using a mixture of the LANL method and the EP method). In addition, its considerably high injection energy is ~2 times of that of the other RFQs in the table, which results in a ~40% larger $\beta$ value for the starting bunching cells. As mentioned in the introduction, the length of a cell for adiabatic bunching is proportional to $\beta^3$, which means that it is more demanding for the HHH RFQ to reach a 3m long structure than the other two RFQs of same length.

In addition, among the 3m long RFQs, only the CERN LINAC4 RFQ and the HHH RFQ have no transverse emittance growth. Noteworthy here is that the transverse input emittance of the HHH RFQ is 20% larger than that of the CERN LINAC4 RFQ and 50% larger than that of the other machines (which means also a larger acceptance). Nevertheless, the longitudinal output emittance of the HHH RFQ is smaller than that of the other two 3m long RFQs. Even at 100mA, the HHH RFQ has still no transverse emittance growth and its output longitudinal emittance is very moderate.

TABLE II. Main design results of some realized ≥300MHz, 3MeV RFQs and the HHH RFQ.

| | SNS | J-PARC RFQ-III | CSNS | KOMAC (PEFP) | CPHS | CERN LINAC4 | HHH 70mA | HHH 100mA |
|---|---|---|---|---|---|---|---|---|
| $\varepsilon_{in, trans., n., rms}$ [π mm mrad] | 0.20 | 0.20 | 0.20 | 0.20 | 0.20 | 0.25 | 0.30 | 0.30 |
| $\varepsilon_{out,x, n., rms}$ [π mm mrad] | 0.21 | 0.21 | 0.2002 | 0.22 | 0.246 | 0.25 | 0.301 | 0.299 |
| $\varepsilon_{out, y, n., rms}$ [π mm mrad] | 0.21 | 0.21 | 0.2002 | 0.22 | 0.248 | 0.25 | 0.297 | 0.299 |
| $\varepsilon_{out, z, rms}$ [keV-deg] | 103 | 110 | 114.3 | 112 | 144 | 130 | 127 | 145 |
| $L$ [m] | 3.7 | 3.6 | 3.6 | 3.2 | 3.0 | 3.0 | 3.0 | 3.0 |
| $T$ [%] | >90 | 98.5 | 97.1 | 98.3 | 97.2 | 95.0 | 96.0 | 90.1 |

## V. CONCLUSIONS

How to avoid the beam instabilities caused by emittance transfer efficiently has been investigated for high current RFQ accelerators. A so-called $\frac{\varepsilon_l}{\varepsilon_t}=1.0$ design guideline is being proposed in order to take advantage of the ideal working area on the Hofmann Charts for safe beam motions at the presence of space charge effects. As the Hofmann Charts are general for accelerator beams, this $\frac{\varepsilon_l}{\varepsilon_t}=1.0$ guideline can be also applied to high current drift tube linac designs.

Compared to the classic RFQ design technique, the LANL Four Section Procedure, which was developed in 1978 – 1980 without taking into account space charge forces [10], the new design approach based on the combination of the $\frac{\varepsilon_l}{\varepsilon_t}=1.0$ design guideline and the practical BABBLE/NFSP method handles the beam bunching more reasonably and more efficiently especially for high current RFQ accelerators.

The beam dynamics simulation results of a 70 – 100mA RFQ chosen for this study have demonstrated that the new design approach can lead to not only high beam transmission and short structure length but also EP-class beam quality without following the EP condition strictly.

## ACKNOWLEDGEMENTS

The author CZ would like to thank Eugene Tanke very much for his friendly and patient help which enabled a deep look inside the design and convenient data analyses as well as his valuable suggestions for the improvement of the manuscript.